\documentclass[onecolumn]{revtex4}
\usepackage{ragged2e}
\usepackage{amsmath}
\usepackage{graphicx}
\usepackage{bm}
\usepackage{soul} 

\usepackage[usenames,dvipsnames]{color}

\definecolor{darkred}{RGB}{0,0,196}

\begin{document}

\title{Extraction of Effective Parameters from Transverse
Momentum Spectra of Heavy Quarkonia in Proton-Proton Collisions at
the LHC} \vspace{0.5cm}

\author{Peng-Cheng~Zhang$^{1,}$\footnote{202312602003@email.sxu.edu.cn},
Hailong~Zhu$^{1,}$\footnote{Correspondence: zhuhl@sxu.edu.cn},
Fu-Hu~Liu$^{1,}$\footnote{Correspondence: fuhuliu@163.com;
fuhuliu@sxu.edu.cn} and
Khusniddin~K.~Olimov$^{2,3,}$\footnote{Correspondence:
khkolimov@gmail.com; kh.olimov@uzsci.net}}

\affiliation{$^1$Institute of Theoretical Physics \& College of
Physics and Electronic Engineering, State Key Laboratory of
Quantum Optics Technologies and Devices \& Collaborative
Innovation Center of Extreme Optics, Shanxi University, Taiyuan
030006, China
\\
$^2$Laboratory of High Energy Physics, Physical-Technical
Institute of Uzbekistan Academy of Sciences, Chingiz Aytmatov Str.
2b, Tashkent 100084, Uzbekistan
\\
$^3$Department of Natural Sciences, National University of Science
and Technology MISIS (NUST MISIS), Almalyk Branch, Almalyk 110105,
Uzbekistan}

\begin{abstract}

\vspace{0.5cm}

\noindent {\bf Abstract:} The effective string tension ($\kappa$)
in the Schwinger mechanism and the effective temperature ($T$) in
Bose-Einstein statistics are extracted from the transverse
momentum ($p_T$) spectra of heavy quarkonia produced in
proton-proton (p+p) collisions at the Large Hadron Collider (LHC).
Here, $T$ derived from the heavy quarkonium $p_T$ spectra also
serves as the initial effective temperature (effective temperature
at the initial stage) of small collision systems. This is because,
despite the absence of quark-gluon plasma (QGP) formation during
the collisions, which leaves $T$ largely unaffected by QGP-related
effects, the initial geometric asymmetry and local partonic
thermalization still induce radial and transverse flows, thereby
contributing to an increase in $T$. The effective parameters
($\kappa$ and $T$) are obtained by fitting the experimental $p_T$
spectra of $J/\psi$ and $\Upsilon(nS)$ ($n=1$, 2, and 3) within
various rapidity intervals, produced in p+p collisions at
center-of-mass energies of $\sqrt{s}=13$ and 8 TeV, as measured by
the LHCb Collaboration. It is found that the multi-component
distribution structured within the framework of the Schwinger
mechanism or Bose-Einstein statistics can effectively describe the
heavy quarkonium $p_T$ spectra in small collision systems. With
decreasing rapidity in the forward region, both $\kappa$ and $T$
increase, indicating a directly proportional relationship between
them. Based on $\kappa$, the average minimum strong force radius
of participant quarks is determined.
\\
\end{abstract}

\maketitle

\parindent=15pt

\section{Introduction}

Current investigations reveal that small collision systems, such
as proton-proton (p+p) or proton-nucleus collisions, also exhibit
collectivity~\cite{1,2,3,4}. This phenomenon suggests that the
temperature concept used in large collision systems, such as
nucleus-nucleus collisions, may be applicable to small
systems~\cite{5,6,7}. However, {there is controversy regarding the
existence of collectivity in small collision systems. If it
exists, collectivity is observed only in high multiplicity events,
which constitute a small fraction (much less than 1\%) of the
total event sample, depending on the collision energy and
experimental selection criteria. In contrast, non-high
multiplicity events, which dominate the event population, do not
exhibit collectivity. Therefore, it is worth discussing whether
the concept of temperature is universally applicable in small
systems.}

Generally, the final temperature of large collision systems can be
extracted from the yields or transverse momentum ($p_T$) spectra
of identified light hadrons, which are produced in thermal
processes during the final kinetic freeze-out stage. In contrast,
the initial temperature of large collision systems can be derived
from heavy quarkonia distribution properties, as these particles
are produced in the early collision stage before fireball
formation~\cite{8,9,10,11}, {if the influence of the increase in
$p_T$ due to radial and transverse flows from the formation and
expansion of quark-gluon plasma (QGP) is removed. To extract a
``real" temperature, the increase in $p_T$ due to radial and
transverse flows from initial geometric asymmetry and local
partonic thermalization should also be removed. Otherwise, the
extracted temperature is an effective temperature.}

Typically, the temperature parameter $T$ used in $p_T$
distributions refers to the so-called effective temperature,
{rather than a direct or indirect measure of the real thermal
temperature of the system. This distinction is necessary} unless
the influence of radial and transverse flows~\cite{12,13,14,15} is
removed from the {$p_T$ spectra. Here, we use the word
``so-called" because there are differences in the definition and
calculation methods of ``effective temperature" in different
fields (such as planetary climate models and non-equilibrium
systems), and it is not an absolute standard concept.}

{To extract effective temperature from particle spectra in
high-energy collisions, applying quantum statistics (Bose-Einstein
and Fermi-Dirac)}~\cite{26,27,28,29,30,31} {is a feasible
approach, though it is not the only method.} Quantum statistics
used to describe particle momentum exhibit remarkable universality
in fitting the $p_T$ spectra of many different particles produced
in large collision systems, although classical (Boltzmann-Gibbs)
statistics serve as a suitable approximation in most cases.

A natural question arises: does this mathematical success also
extend to smaller p+p or proton-nucleus collisions? If so, {is
local thermal equilibrium at particle level or partonic level also
present in small system collisions? Furthermore, is temperature
parameter from quantum statistics applicable for small system
collisions? These related questions have attracted significant
attention from both experimental and theoretical researchers.
Based on the grand canonical ensemble and thermodynamically
consistent non-equilibrium statistics, many researchers argue that
quantum statistics is expected to be successful for small
collision systems. A local partonic thermalization and equilibrium
are achievable in small systems. Therefore, one may attempt to use
a temperature parameter in small systems.}

In Bose-Einstein statistics, $T$ derived from the $p_T$ spectra of
heavy quarkonia in p+p collisions may reflect the initial
effective temperature {(effective temperature at the initial
stage), as flow effects from initial geometric asymmetry and local
partonic thermalization influence the measurement, while these
systems are unlikely to involve QGP formation.} In other words, if
the $p_T$ distribution with the parameter $T$ in Bose-Einstein
statistics is used to describe the production of heavy quarkonia
in p+p collisions~\cite{16,17,18}, $T$ reflects the initial
{effective} temperature of the collisions. Although the initial
{effective} temperature has received relatively little research
attention compared to the extensively studied final temperature
(chemical or kinetic freeze-out temperature), it remains an
important quantity for understanding the early dynamics of
collision systems.

Of course, a single-component distribution from quantum statistics
cannot adequately describe the $p_T$ spectra in small collision
systems. Instead, two- or multi-component distributions play a
significant role. As products generated at the initial stage, the
behavior of heavy quarkonia, such as $J/\psi$ and $\Upsilon(nS)$
($n=1$, 2, and 3), reflects the strength of initial interactions
among participant partons {that are in a state of local thermal
equilibrium consisting of multiple partons.} The multi-component
distribution derived from Bose-Einstein statistics (or the
multi-component Bose-Einstein distribution) can describe the $p_T$
spectra of $J/\psi$ and $\Upsilon(nS)$ and extract the initial
{effective} temperature $T$ of small collision systems.

It is well understood that the mechanism of heavy quarkonium
production in high-energy collisions involves the synergistic
effects of perturbative and non-perturbative quantum
chromodynamics (QCD)~\cite{19,20,21,22,23,24,25}. Despite heavy
quarkonia not being {produced with a high yield in a single
event}, statistical results from a large number of events exhibit
patterns similar to thermal processes. {In addition, heavy
quarkonia are in fact produced in an environment of multiple
partons which are expected to form a state with local
thermalization and then equilibrium, even in p+p collisions.}
Therefore, the statistical laws applicable to thermal processes
can also describe the production of heavy quarkonia.

Due to differing interaction mechanisms {and production stages}
compared to light particles, {$T$ extracted from heavy quarkonium
spectra may differ from that extracted from light particle
spectra. As mentioned above,} we refer to $T$ from heavy
quarkonium spectra in p+p collisions as the initial {effective}
temperature, {because most heavy quarkonia are produced at the
initial stage of the collisions. Even for $J/\psi$ from $b$
quarks, since most $b$ quarks are produced initially, $J/\psi$
from $b$ indirectly carries information about the initial stage.
The fraction of $b$ quarks formed through quark recombination in
the final hadronization stage is negligible to the total $b$
yield, though its precise fraction remains uncertain.}

Although statistical distributions can fit the shape of $p_T$
spectra, {their underlying physical origin remains unclear.} They
do not explain how heavy quarkonia are initially generated, nor do
they connect to fundamental QCD parameters such as string tension,
instead merely describing the phenomenological characteristics of
the final-state distribution {under the condition of local
equilibrium.} To explore the physical origin of heavy quarkonium
production, {we propose that the Schwinger mechanism offers a
natural framework. This mechanism explains how particle pairs are
generated under strong color fields, which is essential for the
initial formation of heavy quarkonia. Here, the effective string
tension $\kappa$ (a parameter distinct from the QCD string
tension) plays a key role: it quantifies the strength of the color
field required to produce these states. Consequently, $\kappa$
directly influences the initial $p_T$ distribution of heavy
quarkonia by determining the energy scale of their generation
process.}

Therefore, to provide a relatively complete description, the
statistical distribution can be combined with the Schwinger
mechanism distribution (or Schwinger distribution). The latter can
describe the interaction strength between two participant partons
in high-energy collisions, though this application remains
relatively uncommon~\cite{32,33,34,35}, even in large collision
systems. Naturally, single-, two-, or multi-component Schwinger
distributions can be employed depending on specific scenarios.
Compared to proton-nucleus and nucleus-nucleus collisions, p+p
collisions offer particular advantages due to the minimal
influence of spectator partons. Here, the participant-spectator
model~\cite{35a,35b,35c,35d}, widely used for proton-nucleus and
nucleus-nucleus collisions, is extended to p+p collisions, where
participants and spectators are considered partons in the study of
heavy quarkonium production.

The Schwinger mechanism describes particle production at the
parton level, while Bose-Einstein statistics characterizes boson
behavior. We aim to extract the effective string tension $\kappa$
from the Schwinger mechanism~\cite{32,33,34,35} and the initial
{effective} temperature $T$ from Bose-Einstein
statistics~\cite{29,30,31}, subsequently investigating the
relationship between {effective parameters} $\kappa$ and $T$. In
this study, multi-component Schwinger and Bose-Einstein
distributions are employed to describe the $p_T$ spectra of
$J/\psi$ and $\Upsilon(nS)$ produced in p+p collisions at
center-of-mass energies $\sqrt{s}=13$ and 8 TeV, as measured by
the LHCb Collaboration at the Large Hadron Collider
(LHC)~\cite{36,37,38,38a}. The {effective} parameters $\kappa$ and
$T$ are extracted, and their relationship is determined.
Additionally, based on $\kappa$, the average minimum strong force
radius is obtained.

The remainder of this article is structured as follows: Section 2
outlines the formalism associated with the Schwinger mechanism and
Bose-Einstein statistics. Results and discussions are presented in
Section 3. Finally, Section 4 provides a summary and conclusions.

\section{Formalism from Schwinger Mechanism and Bose-Einstein Statistics}

Based on the Schwinger mechanism~\cite{32,33,34,35}, a
multi-component distribution can be formulated. Each component $i$
arises from the contribution of two participant partons, where
each parton $j$ contributes transverse momentum ($p_{tj}$) via a
Gaussian-type probability density function $f_{ij}(p_{tj})$. This
is expressed as:
\begin{align}
f_{ij}(p_{tj}) &= C_0(\kappa_i)\exp\bigg[-\frac{\pi(p_{tj}^2+m_0^2)}{\kappa_i}\bigg] \nonumber\\
&= C_0(\kappa_i)\exp\bigg(-\frac{\pi m_0^2}{\kappa_i}\bigg)\exp\bigg(-\frac{\pi p_{tj}^2}{\kappa_i}\bigg) \nonumber\\
&= \frac{1}{\sqrt{\kappa_i}}\exp\bigg(-\frac{\pi
p_{tj}^2}{\kappa_i}\bigg),
\end{align}
where $m_0$ represents the rest mass of the participant parton,
$\kappa_i$ denotes the effective string tension, and
$C_0(\kappa_i)$ is the normalization constant. Without causing
confusion, the parameter $\kappa_i$ can be directly referred to as
string tension for conciseness.

The final $p_T$ of a particle in component $i$ results from the
contributions of two participant partons, following the folding of
their individual contributions. The $p_T$ distribution in
component $i$ is given by:
\begin{align}
f_i(p_T) &= \int_0^{p_T} f_{i1}(p_{t1})f_{i2}(p_T-p_{t1})dp_{t1} \nonumber\\
&=\int_0^{p_T} f_{i2}(p_{t2})f_{i1}(p_T-p_{t2})dp_{t2} \nonumber\\
&=\frac{1}{\kappa_i}\int_0^{p_T}\exp\bigg\{-\frac{\pi\big[p_{t1}^2+(p_T-p_{t1})^2\big]}{\kappa_i}\bigg\}dp_{t1} \nonumber\\
&=\frac{1}{\kappa_i}\int_0^{p_T}\exp\bigg\{-\frac{\pi\big[p_{t2}^2+(p_T-p_{t2})^2\big]}{\kappa_i}\bigg\}dp_{t2},
\end{align}
which is normalized to unity. Let $k_i$ denote the fraction
contributed by component $i$. The final $p_T$ distribution from
the multi-component system can then be written as:
\begin{align}
f(p_T) = \sum_i k_i f_i(p_T),
\end{align}
with $\sum_i k_i = 1$ due to normalization. The average effective
string tension weighted by $k_i$ in the multi-component Schwinger
distribution is $\kappa = \sum_i k_i \kappa_i$.

{In the study of heavy quarkonium production, $\kappa$ in the
Schwinger distribution describes the interaction strength of
participant heavy quarks at the initial stage of the collisions.
Generally, a larger $\kappa$ corresponds to a more violent
collision at higher energy. Correspondingly, the distance between
the two participant heavy quarks becomes shorter, and the parton
number density and energy density become higher. If the Schwinger
distribution is used to fit the spectra of other particles which
are produced at the intermediate and final stages, the
corresponding $\kappa$ is smaller than that at the initial stage.
Parameter variations across different components in the
multi-component Schwinger distribution reflect fluctuations of the
effective string tension.}

In terms of Bose-Einstein statistics~\cite{29,30,31}, for
component $i$ with effective temperature $T_i$, the invariant
yield or boson momentum ($p$) distribution is expressed
as~\cite{29}:
\begin{align}
E\frac{d^3N_i}{d^3p} =
\frac{gV_i}{(2\pi)^3}E\bigg[\exp\bigg(\frac{E-\mu}{T_i}\bigg)-1\bigg]^{-1},
\end{align}
where $N_i$ is the number of bosons, $g=2s+1$ is the degeneracy
factor, $s=1$ is the spin for $J/\psi$ and $\Upsilon(nS)$, $\mu$
is the chemical potential (close to zero at high energy),
$E=\sqrt{p^2+m_0'^2}=m_T\cosh y$ is the energy, $m_0'$ is the rest
mass, $m_T=\sqrt{p_T^2+m_0'^2}$ is the transverse mass,
$y=(1/2)\ln[(E+p_z)/(E-p_z)]$ is the rapidity, $p_z$ is the
longitudinal momentum of the considered boson, and $V_i$ is the
volume of the collision system.

The probability density function of $p_T$ in component $i$ is:
\begin{align}
f_i(p_T) &= \frac{1}{N_i}\frac{dN_i}{dp_T} \nonumber\\
&= \frac{1}{N_i}\frac{gV_i}{(2\pi)^2} p_T \sqrt{p_T^2+m_0'^2} \int_{y_{\min}}^{y_{\max}} \cosh y \nonumber\\
&\quad \times \bigg[\exp\bigg(\frac{\sqrt{p_T^2+m_0'^2}\cosh
y-\mu}{T_i}\bigg)-1\bigg]^{-1}dy,
\end{align}
where $y_{\min}$ and $y_{\max}$ are the minimum and maximum
rapidities, respectively, within the experimental rapidity
interval $[y_{\min},y_{\max}]$. The multi-component distribution
is expressed as:
\begin{align}
f(p_T) = \frac{1}{N}\frac{dN}{dp_T} = \sum_i
k_i\frac{1}{N_i}\frac{dN_i}{dp_T} = \sum_i k_i f_i(p_T),
\end{align}
which is consistent with Eq. (3). The average effective
temperature from the multi-component Bose-Einstein distribution is
given by $T = \sum_i k_i T_i$.

{In the study of heavy quarkonium production, $T$ in the
Bose-Einstein distribution is in fact the initial effective
temperature. This is because the majority of heavy quarkonia are
produced at the initial stage of the collisions, where radial and
transverse flows exist due to the initial geometric asymmetry and
local partonic thermalization. In this context, $T$ serves as an
indicator of the comprehensive outcomes of the system's initial
excitation degree and flow effects. Even so, the larger $T$, the
higher the excitation degree and energy density. Throughout the
entire evolution process of the collision system, the initial
effective temperature $T$ is the highest, and it decreases over
time.}

{In the application of the Bose-Einstein distribution, an
approximate local thermal equilibrium is expected to form in p+p
collisions due to the large number of partons (valence quarks, sea
quarks, and gluons) existing in the interaction system, even if
the yield of specific particles is low. Meanwhile, the statistics
of the experimental sample are very large. One may consider the
frameworks of the grand canonical ensemble and thermodynamically
consistent non-equilibrium statistics. The related statistical
laws describing equilibrium states are available for p+p
collisions. Parameter variations across different components in
the multi-component Bose-Einstein distribution reflect temperature
fluctuations.}

We would like to point out that, when applied to p+p collisions,
both the Schwinger and Bose-Einstein distributions can be
considered as empirical models with effective parameters, which
may have little relevance to thermodynamics. However, these
effective parameters can be useful for heavy-ion physicists, as
they can be compared with those extracted from heavy-ion data. In
the multi-component Schwinger (Bose-Einstein) distribution, the
functional forms of various components remain consistent, despite
differences in parameter values due to observed commonalities,
similarities~\cite{40,41,42,43}, and
universality~\cite{44,45,46,47} in high-energy collisions.
Variations in parameter values arise from differing strengths or
degrees of analogous processes.

\section{Results and Discussion}

\subsection{Comparison with experimental data}

Figure 1 illustrates the double differential cross section,
$d^2\sigma/(dydp_T)$, of (a) prompt $J/\psi$ and (b) $J/\psi$
originating from $b$-quarks in p+p collisions at $\sqrt{s}=13$
TeV, where $\sigma$ denotes the cross section. Figure 2 depicts
the double differential cross section, $d^2\sigma/(dydp_T)$, of
(a) $\Upsilon(1S)$, (b) $\Upsilon(2S)$, and (c) $\Upsilon(3S)$
produced in p+p collisions at $\sqrt{s}=13$ TeV. Similarly, Figure
3 displays the double differential cross section,
$d^2\sigma/(dydp_T)$, of (a) prompt $J/\psi$ and (b) $J/\psi$ from
$b$-quarks in p+p collisions at $\sqrt{s}=8$ TeV, while Figure 4
presents the double differential cross section,
$d^2\sigma/(dydp_T)$, of (a) $\Upsilon(1S)$, (b) $\Upsilon(2S)$,
and (c) $\Upsilon(3S)$ produced in p+p collisions at $\sqrt{s}=8$
TeV. In Figures 1--4, different symbols represent experimental
data measured by the LHCb Collaboration in various $y$ intervals
within the forward rapidity region at the LHC~\cite{36,37,38,38a}.
Solid and dashed curves correspond to our results fitted using
three-component Schwinger and Bose-Einstein distributions,
respectively. To enhance clarity, experimental data and fitted
results are rescaled by factors indicated in the panels. Using the
least square method, optimal parameter values are obtained during
fitting. It is evident that the experimental double differential
cross sections of $J/\psi$ and $\Upsilon(nS)$ across different $y$
intervals in 13 and 8 TeV p+p collisions, as measured by the LHCb
Collaboration, can be accurately described by three-component
Schwinger and Bose-Einstein distributions.
\\

\begin{figure*}[htb!]
\begin{center}
\includegraphics[width=15.5cm]{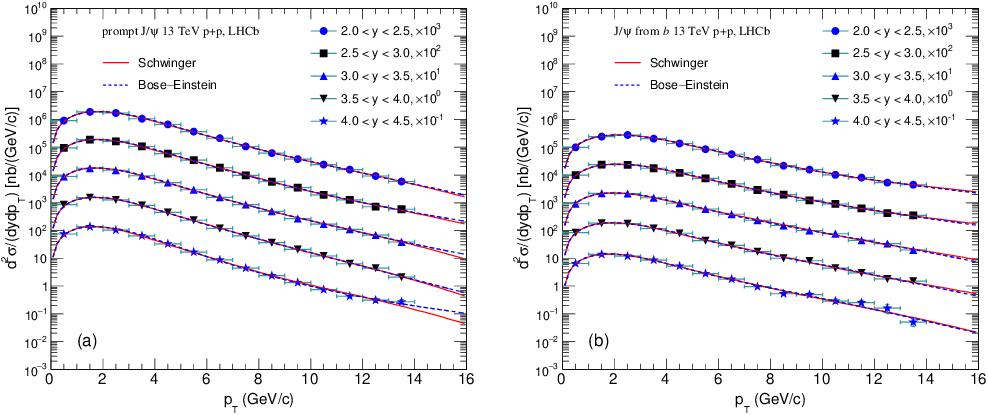}
\end{center}
\justifying\noindent {Figure 1. The double differential cross
section, $d^2\sigma/(dydp_T)$, for (a) prompt $J/\psi$ and (b)
$J/\psi$ from $b$ produced in p+p collisions at $\sqrt{s}=13$ TeV.
Different symbols represent experimental data in various $y$
intervals measured by the LHCb Collaboration~\cite{36}, with the
data rescaled by different factors for clarity. The solid and
dashed curves correspond to our results fitted using
three-component Schwinger and Bose-Einstein distributions,
respectively.}
\end{figure*}

\begin{figure*}[htb!]
\begin{center}
\includegraphics[width=15.5cm]{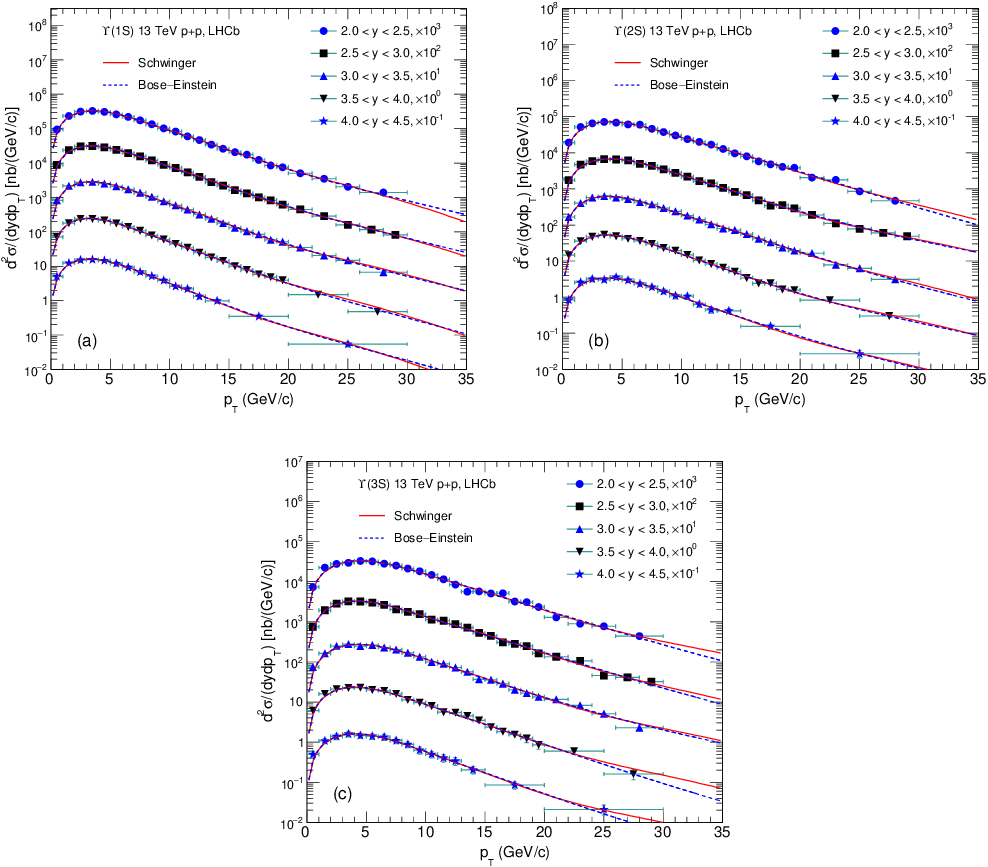}
\end{center}
\justifying\noindent {Figure 2. The double differential cross
section, $d^2\sigma/(dydp_T)$, for (a) $\Upsilon(1S)$, (b)
$\Upsilon(2S)$, and (c) $\Upsilon(3S)$ produced in p+p collisions
at $\sqrt{s}=13$ TeV. Different symbols represent experimental
data in various $y$ intervals measured by the LHCb
Collaboration~\cite{37}, with the data rescaled by different
factors for clarity. The solid and dashed curves correspond to our
results fitted using three-component Schwinger and Bose-Einstein
distributions, respectively.}
\end{figure*}

\begin{figure*}[htb!]
\begin{center}
\includegraphics[width=15.5cm]{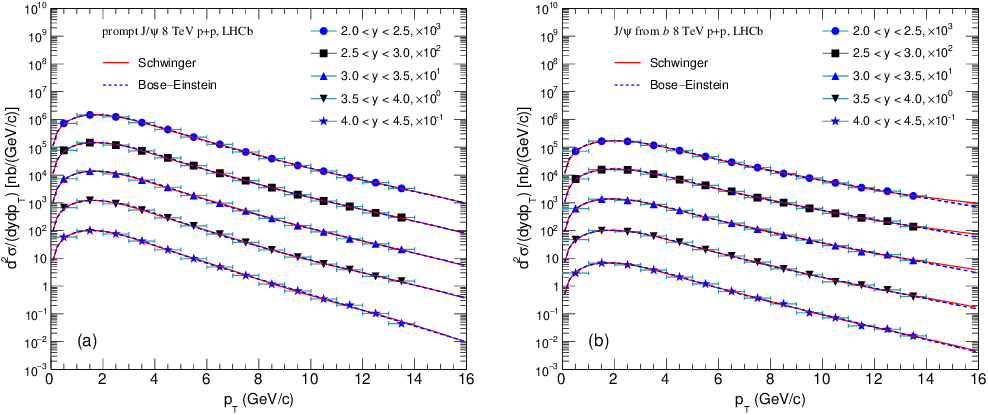}
\end{center}
\justifying\noindent {Figure 3. The double differential cross
section, $d^2\sigma/(dydp_T)$, for (a) prompt $J/\psi$ and (b)
$J/\psi$ from $b$ produced in p+p collisions at $\sqrt{s}=8$ TeV.
Different symbols represent experimental data in various $y$
intervals measured by the LHCb Collaboration~\cite{38}, with the
data rescaled by different factors for clarity. The solid and
dashed curves correspond to our results fitted using
three-component Schwinger and Bose-Einstein distributions,
respectively.}
\end{figure*}

\begin{figure*}[htb!]
\begin{center}
\includegraphics[width=15.5cm]{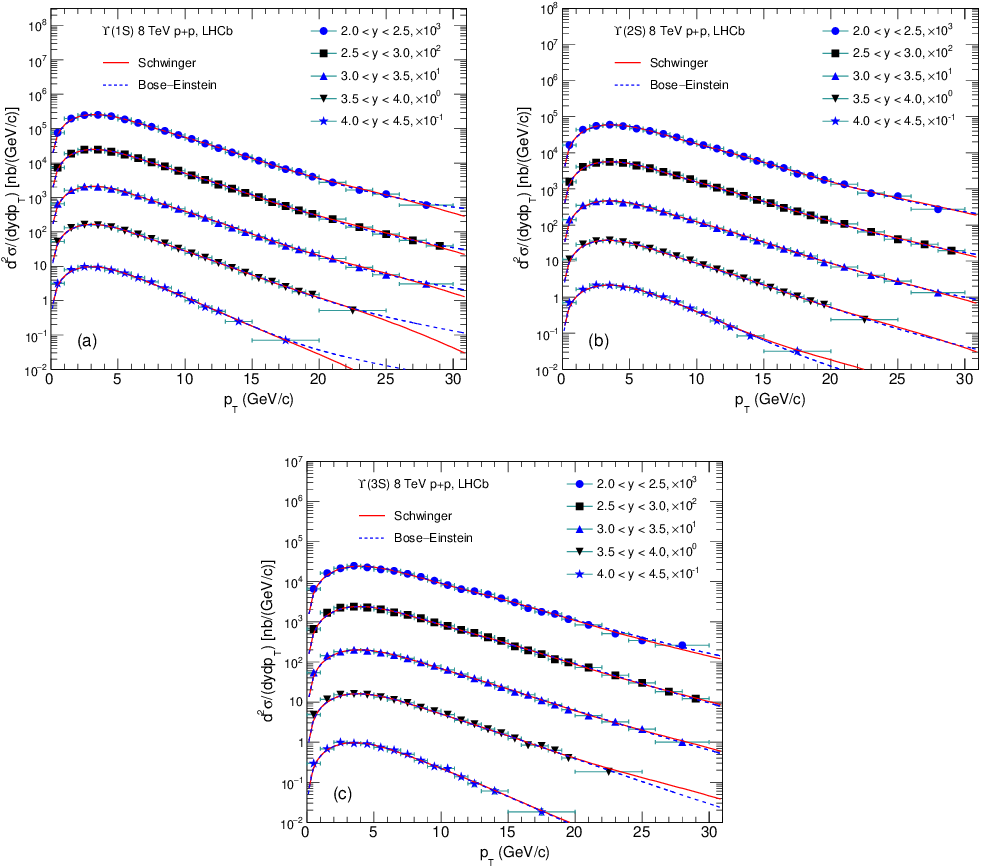}
\end{center}
\justifying\noindent {Figure 4. The double differential cross
section, $d^2\sigma/(dydp_T)$, for (a) $\Upsilon(1S)$, (b)
$\Upsilon(2S)$, and (c) $\Upsilon(3S)$ produced in p+p collisions
at $\sqrt{s}=8$ TeV. Different symbols represent experimental data
in various $y$ intervals measured by the LHCb
Collaboration~\cite{38a}, with the data rescaled by different
factors for clarity. The solid and dashed curves correspond to our
results fitted using three-component Schwinger and Bose-Einstein
distributions, respectively.}
\end{figure*}

\begin{figure*}[htb!]
\begin{center}
\includegraphics[width=15.5cm]{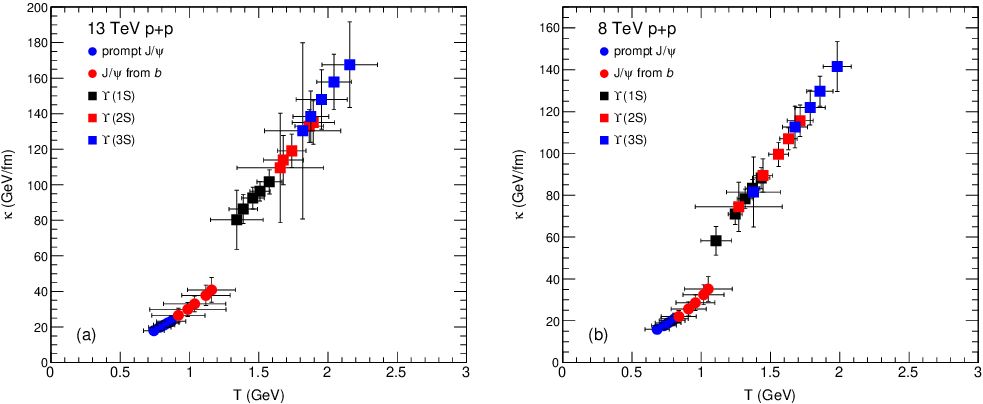}
\end{center}
\justifying\noindent {Figure 5. Relationship between effective
string tension $\kappa$ and initial effective temperature $T$
derived from the spectra of heavy quarkonia $J/\psi$ and
$\Upsilon(nS)$ produced in p+p collisions at (a) 13 TeV and (b) 8
TeV. Results corresponding to $2.0<y<2.5$ ($4.0<y<4.5$) are
located at the higher (lower) value edge for each case. Five cases
are illustrated in the panels.\\}
\end{figure*}

\begin{figure*}[htb!]
\begin{center}
\includegraphics[width=15.5cm]{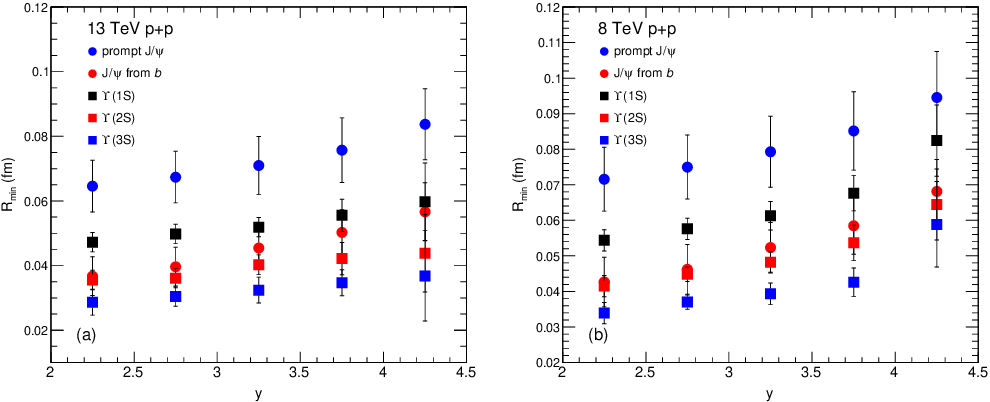}
\end{center}
\justifying\noindent {Figure 6. Dependence of the average minimum
strong force radius $R_{\min}$ of participant $c$ or $b$ quarks on
the rapidity $y$ of quarkonium extracted from p+p collisions at
(a) 13 TeV and (b) 8 TeV. Five cases are illustrated in the
panels.}
\end{figure*}

To explore the relationship between {effective parameters}
$\kappa$ and $T$, Figures 5(a) and 5(b) depict the results derived
from the spectra of heavy quarkonia $J/\psi$ and $\Upsilon(nS)$
produced in p+p collisions at $\sqrt{s}=13$ and 8 TeV,
respectively. For each case
--- i) prompt $J/\psi$, ii) $J/\psi$ from $b$, iii)
$\Upsilon(1S)$, iv) $\Upsilon(2S)$, and v) $\Upsilon(3S)$ --- the
result corresponding to $2.0<y<2.5$ is positioned at the higher
edge, whereas the result for $4.0<y<4.5$ is located at the lower
edge. {The values of $\kappa$ are in the range of $\sim20$--165
GeV/fm at $\sqrt{s}=13$ TeV and $\sim15$--140 GeV/fm at
$\sqrt{s}=8$ TeV. These values are much larger than the classical
string tension value ($\sim 1$ GeV/fm) in QCD vacuum. The values
of $T$ are in the range of $\sim0.75$--2.15 GeV at $\sqrt{s}=13$
TeV and $\sim0.69$--1.98 GeV at $\sqrt{s}=8$ TeV. These values are
indeed very large when compared them to typical QGP freeze-out
temperature ($\sim 0.16$ GeV) in central nucleus-nucleus
collisions at high energy}~\cite{47a,47b}. {Both values of
$\kappa$ and $T$} gradually increase from $4.0<y<4.5$ to
$2.0<y<2.5$. It is observed that the relationship between $\kappa$
and $T$ is approximately linear for each case. However,
considering all five cases collectively, the relationship between
$\kappa$ and $T$ exhibits complexity.

The $\kappa$--$T$ relationship essentially characterizes how the
initial energy of a collision is ``thermalized" (or more broadly,
homogenized) into the efficiency or process of the observed
momentum distribution. This is because $\kappa$ characterizes the
energy density or field strength of the initial collision state
deposited at a local spatial point (the potential barrier that
needs to be overcome to generate quark pairs through the Schwinger
mechanism), while $T$ {reflects the density of these deposited
energies,} which is transformed and redistributed into the degree
of particle transverse motion through the interactions,
scatterings, and possible multi-body effects between some
particles before the system reaches its final observation state.
The positively correlated $\kappa$--$T$ relationship indicates
that the higher the initial energy density, the greater the
density of generated partons, leading to a significant increase in
the chance of secondary scattering and energy exchange between
partons.

{The observed} positive correlation between $\kappa$ and $T$
{suggests} that the system {may exhibit} collective behavior
driven by {high parton} density. In p+p collisions, this may point
to fluid-like characteristics or strong parton cascades that occur
in high multiplicity events. In our study, $\kappa$ is the key
link between the initial state condition and possible parton
collective behavior. The collective effect in small collision
systems is rooted in the inherent multi-body interactions of dense
partons generated during high-energy density events, and $\kappa$
is a measurable proxy variable for this initial energy density,
while $T$ is the macroscopic manifestation of the corresponding
interaction strength. The positive correlation between $\kappa$
and $T$ that we obtained in Figure 5 suggests that a larger
$\kappa$ (indicating higher energy density) should be associated
with more significant parton collectivity. {Due to complexity, our
interpretation on the evidence of parton collective behavior in
p+p collisions remains subject to further empirical validation.}

Averagely, the values of $\kappa$ and $T$ extracted from the
spectra of prompt $J/\psi$, $J/\psi$ from $b$, $\Upsilon(1S)$,
$\Upsilon(2S)$, and $\Upsilon(3S)$ increase in an orderly manner,
though there is no clear demarcation seen in the trend for
$\Upsilon(1S)$, $\Upsilon(2S)$, and $\Upsilon(3S)$ at 8 TeV. In
particular, the slopes of charm and bottom candidates are
different due to their mass difference. This reflects the
interaction strength between two participant partons as well as
the magnitude of the average $p_T$ of the bosons themselves. {Our
results show that} $T$ is significantly larger than the chemical
freeze-out temperature ($\sim0.16$ GeV) determined by the
statistical (thermal) model and hydrodynamical
description~\cite{47a,47b,47c,47d} and the kinetic freeze-out
temperature ($\sim0.10$--0.16 GeV) derived from the blast wave
model~\cite{47e,47f,47g,47h}. {This apparent ``high" value of $T$
arises because it is extracted from the initial stage of the
collision, where the energy density is maximized and the system is
dominated by partonic degrees of freedom. In contrast, chemical
and kinetic freeze-out temperatures are obtained in later stages,
after the system has undergone significant expansion, cooling, and
hadronization, leading to a much lower energy density. The
temperature evolution is also influenced by the transition (if it
exists) from partonic to hadronic matter, which further reduces
the considered temperature. As a result, this relative size
appears to be a natural outcome.}

Although the initial temperatures extracted by us are
significantly higher than typical QGP freeze-out temperature
($\sim0.16$ GeV)~\cite{47a,47b}, our results are comparable to
those obtained using the color string percolation
method~\cite{47i,47j,47k}, which was employed in our previous
work~\cite{47l,47m,47n}; the relative fugacities of quarks and
gluons method~\cite{47o}; a comprehensive heavy ion model
evaluation and reporting algorithm~\cite{47p}; and a perfect
relativistic hydrodynamic solution based on direct photon
observables~\cite{47q,47r,47s,47t}. The present results are lower
than those from the color string percolation method
($\sim2.5$--6.5 GeV)~\cite{47l} at the same energy but higher than
those from the relative fugacity method ($\sim0.2$--0.6 GeV at
$\sqrt{s_{NN}}=200$--5020 GeV)~\cite{47o}, the comprehensive
algorithm ($\sim0.3$--0.4 GeV at $\sqrt{s_{NN}}=200$
GeV)~\cite{47p}, and the perfect hydrodynamic solution
($\sim0.4$--0.5 GeV at $\sqrt{s_{NN}}=200$
GeV)~\cite{47q,47r,47s,47t}, although some of these methods focus
on lower collision energies. In our opinion, if the definitions of
initial, chemical freeze-out, and kinetic freeze-out temperatures
are consistent and well-coordinated, a direct comparison among
them becomes feasible. As the collision system evolves over time,
its temperature gradually decreases.

{It should be noted that while $T$ is extracted from the initial
stage of the collision and does not directly describe the QGP
medium (if it exists) temperature, it indirectly quantifies the
energy scale of the initial state, which is sensitive to medium
effects in small collision systems. Specifically, the value of $T$
reflects the energy density deposited at the collision point,
which can be influenced by the presence of a transient QGP-like
medium or other collective phenomena (if they exist). In small
collision systems, where QGP formation is debated, $T$ serves as a
probe of the initial energy deposition and its subsequent
evolution, which may include medium-induced modifications such as
parton energy loss or collective flow. Thus, by comparing $T$
across different systems or multiplicities, we can infer the
impact of medium effects on the initial state dynamics, even if
$T$ itself is not a temperature of the QGP medium. Generally
speaking, the more obvious the medium effects (or the higher the
multiplicity), the higher the $T$, and the two show a positive
correlation, but there is no causal relationship between the two.}

To understand how the {effective} parameters $\kappa$ and $T$
depend on the collision energy $\sqrt{s}$, we have conducted an
estimation. It is known that $\kappa$ and $T$ are primarily
determined by the energy density of the collision
system~\cite{47u}. As the initial temperature, $T$ is higher than
the medium temperature of QGP (if it exists) due to the fact that
the former occurred earlier than the latter. According to lattice
QCD calculations~\cite{47u}, the energy density is proportional to
the fourth power of the temperature, and it increases with
increasing $\sqrt{s}$. Consequently, both $\kappa$ and $T$ exhibit
a nonlinear growth trend as $\sqrt{s}$ increases. At $\sqrt{s}=13$
and 8 TeV discussed in this work, the ratio of the parameters at
different energies aligns more closely with $\ln(13000)/\ln(8000)$
than with $13000/8000$ or $\sqrt{13000/8000}$, where $\sqrt{s}$ is
expressed in GeV to ensure parameter consistency. This suggests a
potential relationship: $\kappa=a_1+b_1\ln\sqrt{s}$ or
$T=a_2+b_2\ln\sqrt{s}$, where $a_{1,2}$ and $b_{1,2}$ represent
the intercept and slope coefficients, respectively. Determining
the precise dependence of these parameters on collision energy
requires further data fitting across various energies in future.

Based on $\kappa$, the average minimum distance between the two
participant quarks [$c\bar c$ for $J/\psi$ or $b\bar b$ for
$\Upsilon(nS)$] can be calculated. According to the Schwinger
mechanism~\cite{32,33,34,35}, the average minimum distance is
given by $2m_0/\kappa$. This distance also corresponds to twice
the average minimum radius $R_{\min}$ of the strong interaction
force between the two quarks, where $R_{\min}=m_0/\kappa$.
{Generally, $R_{\min}$ represents the closest approach of
participant quarks during the collision, reflecting the initial
overlap degree and energy density. Definitely and physically,
$R_{\min}$ does not represent the physical size of the quarkonia
(which is determined by the bound-state wavefunction), but rather
the characteristic length scale at which the color field is strong
enough to spontaneously produce quark-antiquark pairs. This scale
reflects the initial energy density required for pair production,
which is a key feature of the Schwinger mechanism in the hybrid
model framework.}

{In experimental observation and relevance,} the dependence of
$R_{\min}$ on $y$, extracted from the spectra of heavy quarkonia
produced in p+p collisions at $\sqrt{s}=13$ and 8 TeV, is
illustrated in Figures 6(a) and 6(b), respectively. It is evident
that $R_{\min}$ decreases as $y$ decreases from $4.0<y<4.5$ to
$2.0<y<2.5$ in the forward rapidity region. This is because the
smaller $y$, the smaller the rapidity shift of heavy quarkonium,
the stronger the collisions, and the closer the two quarks are in
the initial state. Extremely small values of $R_{\min}$
($0.03\sim0.09$ fm) are obtained, which are one tenth of various
radii (hard-core, charge, mass, axial) of nucleons ($0.3\sim1.0$
fm)~\cite{48,49,50}. {The small $R_{\min}$ values indicate that
the initial energy density is high enough to produce quark pairs
at very short distances, which is consistent with the expectations
of the Schwinger mechanism in high-energy collisions.}

{On theoretical justification and model relevance,} it should be
noted that the estimation of $R_{\min}$ is not an arbitrary
empirical formula; it comes from the hybrid model framework
discussed in this work, which uses the generation probability of
Schwinger mechanism. When we associate this mechanism with the
dynamics of strings, the minimum energy required to produce a pair
of quarks with mass $m_0$ is related to the work done to stretch a
string to a length of $2R_{\min}$, that is, $2m_0\sim2\kappa
R_{\min}$. From this, we can get the scaling relationship $R_{\rm
min}\sim m_0/\kappa$. Therefore, $R_{\min}$ is a derived quantity
directly calculated from $\kappa$ and $m_0$ obtained by fitting,
and its value reflects the typical spatial scale of producing
heavy quark pairs at a given initial energy density. {The
relevance of this model lies in its ability to connect the initial
energy density (characterized by $\kappa$) to the observable $p_T$
spectra of heavy quarkonia, providing a physical interpretation
for $T$ extracted from the spectra. By studying $R_{\rm min}$, we
gain insights into the initial conditions of the collision, which
are crucial for understanding the production mechanisms of heavy
quarkonia in small collision systems.}

We would like to emphasize that the average minimum strong force
radii (Figure 6) of $c$ and $b$ quarks are significantly smaller
than the sizes ($\sim0.4$ and $\sim0.2$ fm) of different quarkonia
[$J/\psi$ and $\Upsilon(nS)$] obtained via the potential
model~\cite{50a,50b,50c,50d}. This indicates that $c$ and $b$
quarks are extremely close when forming quarkonia. Nevertheless,
within quarkonia, $c$ and $b$ quarks still possess sufficient
space for movement. This situation resembles that of quarks within
nucleons, where quarks are much smaller than the nucleons
themselves. The radius of quarkonium is typically calculated based
on its form factor or decay parameters, including lattice QCD, QCD
sum rules, and related data analyses. The analysis of the average
minimum strong force radius $R_{\min}$ of heavy quarks presented
in this article provides an indirect measurement derived from the
$p_T$ spectra of quarkonia, although $R_{\min}$ does not directly
correspond to the size of quarkonia.

The value of $R_{\min}$ is approximately an order of magnitude
smaller than nucleon radii. Such a small radius is physically
plausible for quark-antiquark binding at the LHC. Due to the
extremely high collision energy, the colliding protons undergo
significant mutual compression, penetration, and overlap, bringing
the quark and antiquark that constitute quarkonium into very close
proximity---potentially even closer than the proton radius. It is
anticipated that at higher collision energies, the value of
$R_{\min}$ will decrease further. Conversely, at lower collision
energies, $R_{\min}$ will increase. If the collision energy is
sufficiently low while still allowing for quarkonium production,
$R_{\min}$ could exceed the proton radius or even be several times
larger than it.

\subsection{General discussion}

{Before continuing with the issues discussed in this article, we
would like to emphasize together the physical meaning and
importance of the extracted parameters ($\kappa$, $T$, $R_{\min}$)
and explain what unique information these parameters provide about
small collision systems and collectivity features. Generally, the
magnitude of $\kappa$ is correlated with the formation of QGP, the
value of $T$ reflects the attainment of local thermal equilibrium,
and the amount of $R_{\min}$ characterizes the geometric
properties of quark interactions. Now, we will elaborate on the
following points.}

{As in the QCD string model, the strength of the interaction
between quark and antiquark pairs can be described by $\kappa$ in
this work. A higher $\kappa$ means stronger binding between quark
and antiquark pairs, which is crucial for understanding the
generation and interaction of heavy quarkonia such as $J/\psi$ and
$\Upsilon(nS)$. By measuring $\kappa$, the non-perturbative
properties of strong interactions in small systems can be
inferred. In small system collisions, changes in $\kappa$ can
reveal the formation and evolution process of QGP: when QGP is
formed, the interaction strength between quark and antiquark pairs
weakens, resulting from a decrease in $\kappa$ value. Therefore,
we observed larger $\kappa$ values, indicating that QGP was not
formed in small system collisions, and these $\kappa$ values have
not been affected by the collectivity of QGP.}

{Similar to the case of large systems, $T$ reflects the thermal
state during the initial collision of small systems and can be
used to characterize the energy density and excitation level of
the system. A higher $T$ means that the system has higher energy
in the early stage of collision, and the measurement of $T$
provides information about energy distribution and thermal
equilibrium state. If a small system is in local thermal
equilibrium, its temperature is higher as a result of localized
energy deposition. The high $T$ values we observed suggest that
the small system may have reached local thermal equilibrium, which
further validates the applicability of the model used.}

{The parameter $R_{\min}$ describes the average minimum
interaction distance between $c$ quark and $\bar c$ (or $b$ quark
and $\bar b$) involved in a collision, reflecting the range of
strong interactions and the geometric properties of
quark-antiquark interactions. The $R_{\min}$ parameter helps
researchers understand the spatial distribution and interaction
range of quarks during collisions: the smaller the $R_{\min}$, the
closer the two quarks are and the shorter the interaction
distance. A smaller $R_{\min}$ may indicate tighter interactions
between quarks, which is crucial for understanding the short-range
properties of strong interactions in small systems. The observed
$R_{\min}$ is much smaller than the nucleon radius, indicating
significant mutual penetration and overlap between the nucleons
involved in the collision, providing experimental evidence for the
short-range nature of strong interactions.}

{Traditionally, research on small system collisions has primarily
focused on macroscopic or bulk observables, such as particle
multiplicity and $p_T$ distributions. In contrast, this study
offers a deeper understanding of small system collisions by
extracting microscopic parameters $\kappa$ and $R_{\min}$, as well
as thermodynamic parameter $T$. The extraction method for these
parameters integrates spectral information from heavy quarkonia,
representing a novel approach that provides direct insights into
strong interactions and thermodynamic states.}

{The overall fitting process not only incorporates the production
mechanisms of heavy quarkonia but also combines modelling
descriptions of thermodynamics and strong interactions. This
comprehensive methodology is innovative in the study of small
system collisions and enables more accurate descriptions of the
complex physical processes involved. Consequently, the fitting
process yields more reliable parameter values, offering new
perspectives and methodological advancements for investigating
small system collisions.}

The dependence of quarkonium production on charged-particle
multiplicity has been extensively studied both experimentally and
theoretically~\cite{50e,50f,50g,50h}. These studies reveal diverse
trends. In general, in high-energy p+p collisions, where
conditions for QGP formation are absent, the quarkonium yield is
mainly influenced by the distribution of initial-state partons and
the final-state hadronization process, with a relatively linear
relationship to the multiplicity of charged particles. In
contrast, in high-energy heavy ion collisions, as the multiplicity
increases, the QGP effect becomes more pronounced, leading to a
nonlinear suppression of quarkonium yield---such as the
``sequential suppression" phenomenon observed in $\Upsilon(nS)$
and $J/\psi$. Heavier collision systems exhibit more significant
suppression. The present work does not investigate the parameter
dependence on charged-particle multiplicity but instead focuses on
quarkonium rapidity, due to differing research objectives.

Comparing relevant parameters across different distributions holds
great significance in high-energy collisions. In our
previous~\cite{51,52,53} and recent studies~\cite{54,55}, we have
investigated the parameters associated with $\kappa$, final
effective temperature, and average $p_T$ extracted from the $p_T$
spectra of particles produced in both large and small collision
systems. Approximate linear correlations under specific
restrictive conditions have been identified. The present work
further validates these observations by comparing $\kappa$ in
Schwinger mechanism with $T$ in Bose-Einstein statistics in p+p
collisions at the LHC. This study provides an important
perspective for researchers to deepen their understanding of
system evolution and interaction mechanisms.

The experimental determination of $\kappa$ can verify the
consistency between the QCD string model and lattice calculation
predictions, providing a benchmark for describing strong
interactions beyond perturbation theory. Simultaneously, when
coupled with non-equilibrium dynamics, $\kappa$ may induce the
formation of strange hadron states (e.g., color superconducting
states). Abnormal fluctuations in the temperature parameter may
indicate the critical region in the QCD phase diagram, verifying
the phase transition paths predicted by lattice QCD. Furthermore,
the correlation between the temperature parameter and boson
occupancy can reveal collective excitations dominated by boson
modes during the QGP de-confinement process.

In the extraction of $\kappa$, there exists a challenge regarding
insufficient real-time detection accuracy in non-perturbative
processes. Future research could focus on developing case
reconstruction algorithms based on deep learning to address this
issue. Similarly, in the extraction of $T$, measurement model
distortion caused by multi-body correlation effects poses a
significant problem. This can potentially be resolved by
constructing thermodynamic evolution equations that incorporate
quantum entanglement. As these are fundamental research tasks,
accurately extracting their values is essential for meaningful
comparisons and understanding their relationships.

Based on the aforementioned discussion, it can be inferred that
through synergistic analysis of $\kappa$ and $T$, researchers can
reconstruct the collision dynamics of the complete spatiotemporal
chain, from the initial energy deposition of the color field to
the evolution of the thermalization medium. This approach allows
for comparing prediction differences between the string model and
the thermal/statistical model for the same observation, enabling
cross-validation of the models. Furthermore, it explores the
possibility of discovering strong interaction corrections or Bose
condensation phases beyond the standard model in the TeV energy
region. In short, the cross-study of these two {effective}
parameters provides a unique perspective for uncovering the deep
structure of matter and the early evolution laws of the universe.

It should be noted that numerous studies have explored light and
heavy flavor production using the Schwinger mechanism combined
with fluctuations in $\kappa$~\cite{67a,67b}. Compared to the high
$p_T$ production of light quarks ($u$, $d$, and $s$), the
production of heavy quarks ($c$ and $b$) shows less suppression,
which is attributed to the dead cone effect~\cite{67c,67d,67e}.
The experimental data cited in this work exhibit a long-tail
distribution, reflecting reduced suppression at high $p_T$, which
is associated with the dead cone effect. Generally, heavy quarks
exhibit a more pronounced dead cone effect due to their large
mass. Gluons, being massless, do not exhibit such an effect, and
the dead cone effect for light quarks is negligible due to their
small mass.

Moreover, if the fluctuations in $\kappa$ follow a Gaussian
distribution, they can lead to thermal production characterized by
exponential $p_T$ distributions. If the fluctuations follow a
Tsallis-type distribution, they result in thermal production at
low $p_T$ and power-law behavior at high $p_T$~\cite{67f,67g}. In
most cases, the observed data align better with a combination of
thermal production at low $p_T$ and power-like production at high
$p_T$. This distinction across $p_T$ regions implies different
underlying interaction mechanisms or degrees of excitation. The
present work employs the same functional form with varying
parameters for different $p_T$ ranges, indicating a
multi-component distribution and resulting in fluctuations in $T$.
These fluctuations also reflect variations in $\kappa$,
highlighting the intrinsic connection between $\kappa$ and $T$.

In the model presented in this work, the parameter $\kappa$
derived from Schwinger mechanism appears in the analytical
expression describing the $p_T$ distribution of particles. It does
not directly measure the linear coefficient of static
quark-antiquark potential. More precisely, $\kappa$ is extracted
by broadening the $p_T$ of heavy quarkonia and associating it with
an effective one-dimensional string or color flow tube model. The
primary significance of extracting $\kappa$ from p+p collisions
lies in its role as a phenomenological scale for describing the
broadening of $p_T$. The long-term goal of this study is to
explore its relationship with the vacuum QCD string tension and
its evolution across different systems (p+p, proton-nucleus, and
nucleus-nucleus).

As the quantity that reflects the intensity or energy density of
the color field generated in the initial state of high-energy
collisions, the parameter $\kappa$ may also correspond to the
square of the average transverse force or momentum transfer
experienced by some partons passing through the color field
region, in the context of multiple scatterings. {The $\kappa$
values obtained by us are much larger than the classical string
tension value ($\sim 1$ GeV/fm) in QCD vacuum.} This may also
suggest that the structure or intensity of the color field differs
from that in vacuum under high-energy-density, local-equilibrium
initial-state conditions. This is a very interesting possibility
that may provide clues for understanding the chromodynamics of
early collision moments.

The possible physical origins associated with this $T$ in small
collision systems can be summarized as follows: (1) The magnitudes
of $p_T$ generated during the initial state string fragmentation
or parton cascade process; (2) The average transverse excitation
energy scale generated by the collective effect of multiple
colored strings in the context of multiple parton interactions;
(3) Due to local equilibrium and high statistics, the local phase
space density may approximate a certain statistical distribution,
and $T$ can be regarded as a measure of it. {In our view, all
three of these origins are valid in p+p collisions.}

\subsection{Further discussion}

Before summarizing and concluding, we would like to emphasize
recent observations of possible collective effects in small
collision systems at the LHC~\cite{68,69}, which also suggest
transient thermalization. {These observations support the local
thermal equilibrium in p+p collisions. In addition,} apart from
heavy quarkonium production, a vast majority of particles produced
in high-energy collisions contain light quarks. It is estimated
that pions account for approximately 80--90\% of the total hadron
yield, while heavy quarkonia contribute less than
1\%~\cite{36,37,38,38a}, with the remaining fraction distributed
among various other particles. {This indicates that the total
number of particles generated in p+p collisions at the LHC is a
considerable amount, which also tends to support the local thermal
equilibrium.}

Although heavy quarks are indeed produced through hard processes,
our focus lies in their hadronization outcomes, where
non-perturbative effects---such as string
fragmentation---dominate. Rather than pursuing detailed
micro-processes, we treat our methods as alternative analysis
tools to extract meaningful quantities that may serve as
references for future studies. If heavy quarks ($c$ and $b$)
produced via perturbative QCD---such as those predicted by Fixed
Order Next-to-Leading Log (FONLL) calculations~\cite{70,71}---are
considered within a microscopic framework, the present work
focuses on the macroscopic behavior of heavy quarkonia.
Specifically, using simple analytical expressions, we perform an
indirect estimation of $\kappa$. Although the extracted value
appears unusually large, it may be refined through further
calibration.

As is known~\cite{54,55}, classical and quantum
statistics~\cite{26,27,28,29,30,31} yield higher $T$ estimates
compared to Tsallis statistics~\cite{72,73,74,75}, which in turn
yields higher values than q-dual statistics~\cite{76}. To unify
these different measures of $T$, one can systematically compare
them and investigate the relationships among various models. Just
as $T$ is model-dependent, it is possible that $\kappa$ is also
model-dependent. To reconcile $\kappa$ used in the Schwinger
mechanism~\cite{32,33,34,35} with that in PYTHIA and other
microscopic models~\cite{77,78,79,80}, one may consider reducing
the amplitude of $\kappa$ in the Schwinger mechanism through an
appropriate correction method, such as scaling it down by a factor
of 1/100 or taking the square root of its original value. The
exact functional form or the relationship between $\kappa$ in the
Schwinger mechanism and other models needs to be determined
through systematic comparisons in future studies.

Some researchers think that the Schwinger mechanism is only
applicable to light particles, possibly due to the narrow $p_T$
spectra of both. For heavy quarkonium, the $p_T$ spectrum is broad
and requires a multi-component distribution with different string
tension parameters, which is understandable. In the narrow $p_T$
region, a small string tension is applicable; while in the wide
$p_T$ region, a large string tension is applicable. In most cases,
a three-component distribution is needed, in which the first,
second, and third components correspond to low, intermediate, and
high $p_T$ regions, respectively. Both the multi-component
Schwinger and Bose-Einstein distributions can be described using
Tsallis statistics with fewer components, which reflects that
Tsallis statistics can smooth out the $\kappa$ and $T$
fluctuations~\cite{51,52} present in multi-component Schwinger and
Bose-Einstein distributions, respectively~\cite{53,54,55}.

According to our comparative analysis, $\kappa$ in the Schwinger
mechanism can be also served as an effective measure of the
initial-stage temperature. Unlike dynamic models, there is no time
evolution associated with the Schwinger string, since its tension
is extracted based on static $p_T$ spectra of specific particles.
This represents a fundamental difference from the string tension
used in the Cornell potential and that obtained from
finite-temperature lattice QCD simulations~\cite{81,82}. If the
Schwinger mechanism is applied to the spectra of various
particles---each produced at different stages of the
collision---it may be possible to extract a time-evolving picture
of $\kappa$. This will be one of the key focuses of our future
research.

\section{Summary and Conclusions}

The transverse momentum spectra (double differential cross
sections) of $J/\psi$ and $\Upsilon(nS)$ produced in p+p
collisions at $\sqrt{s}=13$ and 8 TeV within different rapidity
intervals in the forward rapidity region are analyzed using the
Schwinger mechanism and Bose-Einstein statistics. The
multi-component distribution successfully fits the experimental
results measured at the LHC by the LHCb Collaboration. Our study
demonstrates that both the effective string tension $\kappa$ and
initial effective temperature $T$ increase as rapidity decreases
in the considered rapidity region, indicating a positive
correlation between {the two effective} parameters.

{The values of $\kappa$ from the Schwinger mechanism are in the
range of $\sim20$--165 GeV/fm at $\sqrt{s}=13$ TeV and
$\sim15$--140 GeV/fm at $\sqrt{s}=8$ TeV. These values are much
larger than the classical string tension value ($\sim 1$ GeV/fm)
in QCD vacuum, indicting that QGP has not formed in small system.
The values of $T$ from the Bose-Einstein statistics are in the
range of $\sim0.75$--2.15 GeV at $\sqrt{s}=13$ TeV and
$\sim0.69$--1.98 GeV at $\sqrt{s}=8$ TeV. These temperature values
are indeed very large when compared them to other temperatures in
high-energy collisions, indicting local thermal equilibrium in
small system and reasonable application of the related laws.}

At higher collision energies, both {the effective parameters}
$\kappa$ and $T$ correspond to larger values, reflecting greater
deposited energy in the collisions. For the production of i)
prompt $J/\psi$, ii) $J/\psi$ from $b$, iii) $\Upsilon(1S)$, iv)
$\Upsilon(2S)$, and v) $\Upsilon(3S)$, both {the effective}
parameters increase sequentially. This suggests that more
collision energy is required for $\Upsilon(nS)$ production due to
the larger mass of participating quarks ($b\bar b$).

Based on the Schwinger mechanism, the average minimum strong force
radii $R_{\min}$ of $c$ and $b$ quarks in the formation of
$J/\psi$ and $\Upsilon(nS)$ are approximately $0.03\sim0.09$ fm.
These values are much smaller than the proton size {(typically
$0.3\sim1.0$ fm), indicating that during the initial stage of
high-energy p+p collisions at the LHC, the two incoming protons
can approach each other so closely that their internal quark and
gluon constituents overlap spatially.}

{The physical meaning of this overlap phenomenon lies in the fact
that, at sufficiently high collision energies, the protons are not
simply hard spheres that scatter elastically; instead, they can
mutually penetrate to distances far smaller than their nominal
radii. This deep overlap creates an extremely high density region
where the color fields of the two protons superpose, enabling the
Schwinger mechanism to produce heavy quark pairs (e.g.,
$c\bar{c}$, $b\bar{b}$) at very short length scales (as
characterized by $R_{\min}$).}

{This study has made significant contributions to the study of
small system collisions by extracting parameters such as $\kappa$,
$T$, and $R_{\min}$. These parameters provide unique information
about strong interactions, thermodynamic states, and the geometry
of quark-antiquark interactions. Meanwhile, the parameter
extraction method and overall fitting process are novel, and these
approaches are of great significance in understanding collective
behavior (if any) and state of matter under extreme conditions in
small systems.}
\\
\\
{\bf Data Availability Statement}
\\
The data used to support the findings of this study and some
outcomes or conclusive statements are included within the article
and are cited at relevant places within the text as references.
\\
\\
{\bf Ethical Statement}
\\
The authors declare that they are in compliance with ethical
standards regarding the content of this paper.
\\
\\
{\bf Disclosure}
\\
The funding agencies have no role in the design of the study; in
the collection, analysis, or interpretation of the data; in the
writing of the manuscript; or in the decision to publish the
results.
\\
\\
{\bf Conflicts of Interest}
\\
The authors declare no conflicts of interest.
\\
\\
{\bf Funding}
\\
The work of Shanxi group was supported by National Natural Science
Foundation of China under Grant No. 12147215, Fundamental Research
Program of Shanxi Province under Grant No. 202303021221071, Shanxi
Scholarship Council of China under Grant Nos. 2023-033 and
2022-014, and the Fund for Shanxi ``1331 Project" Key Subjects
Construction. The work of K.K.O. was supported by the Agency of
Innovative Development under the Ministry of Higher Education,
Science and Innovations of the Republic of Uzbekistan within the
fundamental project No. F3-20200929146 on analysis of open data on
heavy-ion collisions at RHIC and LHC.
\\
\\

{\small
\begin{thebibliography}{99}

\setlength{\itemsep}{2pt}

\bibitem{1}
S. Guo, H. S. Wang, K. Zhou, and G. L. Ma, ``Machine learning
study to identify collective flow in small and large colliding
systems," {\it Physical Review C} 110, no. 2 (2024): 024910,
https://doi.org/10.1103/PhysRevC.110.024910.

\bibitem{2}
R. S. Bhalerao, ``Collectivity in large and small systems formed
in ultrarelativistic collisions," {\it The European Physical
Journal Special Topics} 230, no. 3 (2021): 635--654,
https://doi.org/10.1140/epjs/s11734-021-00019-x.

\bibitem{3}
R. A. Lacey (for the STAR Collaboration), ``Long-range
collectivity in small collision-systems with two- and
four-particle correlations @ STAR," {\it Nuclear Physics A} 1005
(2021): 122041, https://doi.org/10.1016/j.nuclphysa.2020.122041.

\bibitem{4}
S. Schlichting and P. Tribedy, ``Collectivity in small collision
systems: an initial-state perspective," {\it Advances in High
Energy Physics} 2016 (2016): 8460349,
http://dx.doi.org/10.1155/2016/8460349.

\bibitem{5}
T. S. Bir{\'o}, {\it Is there a temperature? conceptual challenges
at high energy, acceleration and complexity} (Spring New York, New
York, USA) (2011).

\bibitem{6}
M. Badshah, M. Waqas, A. M. Khubrani, and M. Ajaz, ``Systematic
analysis of the pp collisions at LHC energies with Tsallis
function," {\it Europhysics Letters} 141, no. 6 (2023): 64002,
http://dx.doi.org/10.1209/0295-5075/acbf6d.

\bibitem{7}
M. Waqas, G. X. Peng, M. Ajaz, A. H. Ismail, Z. Wazir, and L. L.
Li, ``Extraction of different temperatures and kinetic freeze-out
volume in high energy collisions," {\it Journal of Physics G} 49,
no. 9 (2022): 095102, http://dx.doi.org/10.1088/1361-6471/ac6a00.

\bibitem{8}
I. M. Lofnes (for the ALICE Collaboration), ``Quarkonia as probes
of the QGP and the initial stages of the heavy-ion collisions with
ALICE," {\it EPJ Web of Conference} 259 (2022): 12004,
https://doi.org/10.1051/epjconf/202225912004.

\bibitem{9}
G. Wolschin, ``Aspects of relativistic heavy-ion collisions," {\it
Universe} {\bf 6}, no. 5 (2020): 61,
https://doi.org/10.3390/universe6050061.

\bibitem{10}
T. Song, C. M. Ko, and S. H. Lee, ``Quarkonium formation time in
relativistic heavy-ion collisions," {\it Physical Review C} 91,
no. 4 (2015): 044909, https://doi.org/10.1103/PhysRevC.91.044909.

\bibitem{11}
P. Khan (for the ALICE Collaboration), ``Upsilon production in
Pb-Pb and p-Pb collisions at forward rapidity with ALICE at the
LHC," {\it Journal of Physics: Conference Series} 509, no. 1
(2014): 012112, https://doi.org/10.1088/1742-6596/509/1/012112.

\bibitem{12}
S. Pandey, S. K. Tiwari, and B. K. Singh, Pseudorapidity density,
transverse momentum spectra, and elliptic flow studies in Xe-Xe
collision systems at $\sqrt{s_{NN}}=5.44$ TeV using the HYDJET++
model," {\it Physical Review C} 103, no. 1 (2021): 014903,
https://doi.org/10.1103/PhysRevC.103.014903.

\bibitem{13}
S. R. Nayak, S. Pandey, and B. K. Singh, ``Beam energy dependence
of transverse momentum distribution and elliptic flow in Au-Au
collisions using HYDJET++ model," {\it The European Physical
Journal Plus} 140, no. 5 (2025): 375,
https://doi.org/10.1140/epjp/s13360-025-06298-w.

\bibitem{14}
M. Badshah, M. Ajaz, M. Waqas, and H. Younis, ``Evolution of
effective temperature, kinetic freeze-out temperature and
transverse flow velocity in pp collision," {\it Physica Scripta}
98, no. 11 (2023): 115306,
https://doi.org/10.1088/1402-4896/ad00eb.

\bibitem{15}
A. M. K. Radhakrishnan, S. Prasad, S. Tripathy, N. Mallick, and R.
Sahoo, ``Investigating radial flow-like effects via pseudorapidity
and transverse spherocity dependence of particle production in pp
collisions at the LHC," {\it The European Physical Journal Plus}
140, no. 2 (2025): 110,
https://doi.org/10.1140/epjp/s13360-025-05996-9.

\bibitem{26}
V. N. Kolokoltsov, ``On a probabilistic derivation of the basic
particle statistics (Bose-Einstein, Fermi-Dirac, canonical,
grand-canonical, intermediate) and related distributions," {\it
Trudy Moskovskogo Matematicheskogo Obshchestva} 82, no. 1 (2021):
93--104 or {\it Transactions of the Moscow Mathematical Society}
82, no. 1 (2021): 77--87, https://doi.org/10.1090/mosc/316.

\bibitem{27}
W. S. Dai and M. Xie, ``Do bosons obey Bose-Einstein distribution:
two iterated limits of Gentile distribution," {\it Physics Letters
A} 373, no. 17 (2009): 1524--1526,
https://doi.org/10.1016/j.physleta.2009.02.054.

\bibitem{28}
H. Hasegawa, ``Bose-Einstein and Fermi-Dirac distributions in
nonextensive quantum statistics: exact and interpolation
approaches," {\it Physical Review E} 80, no. 1 (2009): 011126,
https://doi.org/10.1103/PhysRevE.80.011126.

\bibitem{29}
J. Cleymans and D. Worku, ``Relativistic thermodynamics:
transverse momentum distributions in high-energy physics," {\it
The European Physical Journal A} 48, no. 11 (2012): 160,
https://doi.org/10.1140/epja/i2012-12160-0.

\bibitem{30}
R. Gupta and S. Jena, ``Model comparison of the transverse
momentum spectra of charged hadrons produced in PbPb collision at
$\sqrt{s_{NN}}=5.02$ TeV," {\it Advances in High Energy Physics}
2022 (2022): 5482034, https://doi.org/10.1155/2022/5482034.

\bibitem{31}
A. Jaiswal and V. Roy, ``Relativistic hydrodynamics in heavy-ion
collisions: general aspects and recent developments," {\it
Advances in High Energy Physics} 2016 (2016): 9623034,
https://doi.org/10.1155/2016/9623034.

\bibitem{16}
J. Hong and S. H. Lee, ``Energy loss of heavy quarkonia in hot QCD
plasmas," {\it Physical Review C} 103, no. 5 (2021): 054907,
https://doi.org/10.1103/PhysRevC.103.054907.

\bibitem{17}
J. M. Durham (for the PHENIX Collaboration), ``Recent quarkonia
studies from the PHENIX experiment," {\it Nuclear Physics A} 982
(2019): 719--722, https://doi.org/10.1016/j.nuclphysa.2018.09.026.

\bibitem{18}
B. Krouppa, A. Rothkopf, and M. Strickland, ``Bottomonium
suppression at RHIC and LHC," {\it Nuclear Physics A} 982 (2019):
727--730, https://doi.org/10.1016/j.nuclphysa.2018.09.034.

\bibitem{19}
R. Sharma and I. Vitev, ``High transverse momentum quarkonium
production and dissociation in heavy ion collisions," {\it
Physical Review C} 87, no. 4 (2013): 044905,
https://doi.org/10.1103/PhysRevC.87.044905.

\bibitem{20}
M. Shifman, ``Persistent challenges of quantum chromodynamics,"
{\it International Journal of Modern Physics A} 21, no. 28n29
(2006): 5695--5720, https://doi.org/10.1142/S0217751X06034914.

\bibitem{21}
A. Mocsy, P. Petreczky, and M. Strickland, ``Quarkonia in the
quark gluon plasma," {\it International Journal of Modern Physics
A} 28, no. 11 (2013): 1340012,
https://doi.org/10.1142/S0217751X13400125.

\bibitem{22}
D. Bala, S. Ali, O. Kaczmarek, Pavan, and HotQCD Collaboration,
``Finite temperature quarkonia spectral functions in the
pseudoscalar channel," {\it Journal of Subatomic Particles and
Cosmology} 3 (2025): 100042,
https://doi.org/10.1016/j.jspc.2025.100042.

\bibitem{23}
A. Vairo, ``Quarkonium dissociation in a thermal bath," {\it AIP
Conference Proceedings} 1701, no. 1 (2016): 020017,
https://doi.org/10.1063/1.4938606.

\bibitem{24}
K. B. Fadafan and S. K. Tabatabaei, ``Thermal width of quarkonium
from holography," {\it The European Physical Journal C} 74, no. 4
(2014): 2842, https://doi.org/10.1140/epjc/s10052-014-2842-2.

\bibitem{25}
N. Brambilla, M. A. Escobedo, J. Ghiglieri, and A. Vairo,
``Thermal width and quarkonium dissociation by inelastic parton
scattering," {\it Journal of High Energy Physics} 2013, no. 5
(2013): 130, https://doi.org/10.1007/JHEP05(2013)130.

\bibitem{32}
J. Schwinger, ``On gauge invariance and vacuum polarization," {\it
Physical Review} 82, no. 5 (1951): 664--679,
https://doi.org/10.1103/PhysRev.82.664.

\bibitem{33}
R. C. Wang and C. Y. Wong, ``Finite-size effect in the Schwinger
particle-production mechanism," {\it Physical Review D} 38, no. 7
(1988): 348--359, https://doi.org/10.1103/PhysRevD.38.348.

\bibitem{34}
P. Braun-Munzinger, K. Redlich, and J. Stachel, ``Particle
production in heavy ion collisions," in: {\it Quark-Gluon Plasma
3}, eds. R. C. Hwa and X. N. Wang (World Scientific, Singapore)
(2004): 491--599.

\bibitem{35}
C. Y. Wong, {\it Introduction to High Energy Heavy Ion Collisions}
(World Scientific, Singapore) (1994).

\bibitem{35a}
R. L. Glauber, in: {\it Lectures in Theoretical Physics}, eds. W.
E. Brittin and L. G. Dunham (Interscience, New York, USA) (1959).

\bibitem{35b}
L. Shi, P. Danielewicz, and R. Lacey, ``Spectator response to the
participant blast," {\it Physical Review C} 64, no. 3 (2001):
034601, https://doi.org/10.1103/PhysRevC.64.034601.

\bibitem{35c}
T. Gaitanos, H. H. Wolter, and C. Fuchs, ``Spectator and
participant decay in heavy ion collisions," {\it Physics Letters
B} 478, nos. 1--3 (2000): 79--85,
https://doi.org/10.1016/S0370-2693(00)00300-2.

\bibitem{35d}
A. D. Sood and R. K. Puri, ``The study of participant-spectator
matter and collision dynamics in heavy-ion collisions," {\it
International Journal of Modern Physics E} 15, no. 4 (2006):
899--910, https://doi.org/10.1142/S0218301306004685.

\bibitem{36}
LHCb Collaboration (R. Aaij {\it et al.}), ``Measurement of
forward $J/\psi$ production cross-sections in $pp$ collisions at
$\sqrt{s}=13$ TeV," {\it Journal of High Energy Physics} 2015, no.
10 (2015): 172, https://doi.org/10.1007/JHEP10(2015)172.

\bibitem{37}
LHCb Collaboration (R. Aaij {\it et al.}), ``Measurement of
$\Upsilon$ production in $pp$ collisions at $\sqrt{s}=13$ TeV,"
{\it Journal of High Energy Physics} 2018, no. 7 (2018): 134,
https://doi.org/10.1007/JHEP07(2018)134.

\bibitem{38}
LHCb Collaboration (R. Aaij {\it et al.}), ``Production of
$J/\psi$ and $\Upsilon$ mesons in $pp$ collisions at $\sqrt{s}=8$
TeV," {\it Journal of High Energy Physics} 2013, no. 6 (2013): 64,
https://doi.org/10.1007/JHEP06(2013)064.

\bibitem{38a}
LHCb Collaboration (R. Aaij {\it et al.}), ``Forward production of
$\Upsilon$ mesons in $pp$ collisions at $\sqrt{s}=7$ and 8 TeV,"
{\it Journal of High Energy Physics} 2015, no. 11 (2015): 103,
https://doi.org/10.1007/JHEP11(2015)103.

\bibitem{40}
A. N. Mishra, A. Ortiz, and G. Paic, ``Intriguing similarities of
high-$p_T$ particle production between $pp$ and $A$-$A$
collisions," {\it Physical Review C} 99, no. 3 (2019): 034911,
https://doi.org/10.1103/PhysRevC.99.034911.

\bibitem{41}
E. K. G. Sarkisyan and A. S. Sakharov, ``Multihadron production
features in different reactions," {\it AIP Conference Proceedings}
828, no. 1 (2006): 35--41, https://doi.org/10.1063/1.2197392.

\bibitem{42}
A. N. Mishra, R. Sahoo, E. K. G. Sarkisyan, and A. S. Sakharov,
``Effective-energy budget in multiparticle production in nuclear
collisions," {\it The European Physical Journal C} 74, no. 11
(2014): 3147, https://doi.org/10.1140/epjc/s10052-014-3147-1 and
``Erratum to: Effective-energy budget in multiparticle production
in nuclear collisions," {\it The European Physical Journal C} 75,
no. 2 (2015): 70, https://doi.org/10.1140/epjc/s10052-015-3275-2.

\bibitem{43}
E. K. G. Sarkisyan and A. S. Sakharov, ``Relating multihadron
production in hadronic and nuclear collisions," {\it The European
Physical Journal C} 70, no. 3 (2010): 533--541,
https://doi.org/10.1140/epjc/s10052-010-1493-1.

\bibitem{44}
E. K. G. Sarkisyan, A. N. Mishra, R. Sahoo, and A. S. Sakharov,
``Multihadron production dynamics exploring the energy balance in
hadronic and nuclear collisions," {\it Physical Review D} 93, no.
5 (2016): 054046, https://doi.org/10.1103/PhysRevD.93.054046 and
``Publisher's note: Multihadron production dynamics exploring the
energy balance in hadronic and nuclear collisions [Phys. Rev. D
93, 054046 (2016)]," {\it Physical Review D} 93, no. 7 (2016):
079904, https://doi.org/10.1103/PhysRevD.93.079904.

\bibitem{45}
E. K. G. Sarkisyan, A. N. Mishra, R. Sahoo, and A. S. Sakharov,
``Centrality dependence of midrapidity density from GeV to TeV
heavy-ion collisions in the effective-energy universality picture
of hadroproduction," {\it Physical Review D} 94, no. 1 (2016):
011501(R), https://doi.org/10.1103/PhysRevD.94.011501.

\bibitem{46}
E. K. G. Sarkisyan, A. N. Mishra, R. Sahoo, and A. S. Sakharov,
``Effective-energy universality approach describing total
multiplicity centrality dependence in heavy-ion collisions," {\it
Europhysics Letters} 127, no. 6 (2019): 62001,
https://doi.org/10.1209/0295-5075/127/62001.

\bibitem{47}
P. Castorina, A. Iorio, D. Lanteri, H. Satz, and M. Spousta,
``Universality in hadronic and nuclear collisions at high energy,"
{\it Physical Review C} 101, no. 5 (2020): 054902,
https://doi.org/10.1103/PhysRevC.101.054902.

\bibitem{47a}
P. Braun-Munzinger, J. Stachel, and C. Wetterich, ``Chemical
freeze-out and the QCD phase transition temperature," {\it Physics
Letters B} 596, nos. 1--2 (2004): 61--69,
https://doi.org/10.1016/j.physletb.2004.05.081.

\bibitem{47b}
A. Andronic, P. Braun-Munzinger, K. Redlich, and J. Stachel,
``Decoding the phase structure of QCD via particle production at
high energy," {\it Nature} 561, no. 7723 (2018): 321--330,
https://doi.org/10.1038/s41586-018-0491-6.

\bibitem{47c}
F. A. Flor, G. Olinger, and R. Bellwied, ``Flavour and energy
dependence of chemical freeze-out temperatures in relativistic
heavy ion collisions from RHIC-BES to LHC energies," {\it Physics
Letters B} 814 (2021): 136098,
https://doi.org/10.1016/j.physletb.2021.136098.

\bibitem{47d}
P. Huovinen, ``Chemical freeze-out temperature in hydrodynamical
description of Au+Au collisions at $\sqrt{s_{NN}}=200$ GeV," {\it
The European Physical Journal A} 37, no. 1 (2008): 121--128,
https://doi.org/10.1140/epja/i2007-10611-3.

\bibitem{47e}
E. Schnedermann, J. Sollfrank, and U. Heinz, ``Thermal
phenomenology of hadrons from 200A GeV S+S collisions," {\it
Physical Review C} 48, no. 5 (1993): 2462--2475,
https://doi.org/10.1103/PhysRevC.48.2462.

\bibitem{47f}
STAR Collaboration (B. I. Abelev {\it et al.}), ``Systematic
measurements of identified particle spectra in pp, d+Au, and Au+Au
collisions at the STAR detector," {\it Physical Review C} 79, no.
3 (2009): 034909, https://doi.org/10.1103/PhysRevC.79.034909.

\bibitem{47g}
STAR Collaboration (B. I. Abelev {\it et al.}), ``Identified
particle production, azimuthal anisotropy, and interferometry
measurements in Au+Au collisions at $\sqrt{s_{NN}}=9.2$ GeV," {\it
Physical Review C} 81, no. 2 (2010): 024911,
https://doi.org/10.1103/PhysRevC.81.024911.

\bibitem{47h}
Z. B. Tang, Y. C. Xu, L. J. Ruan, G. van Buren, F. Q. Wang, and Z.
B. Xu, ``Spectra and radial flow in relativistic heavy ion
collisions with Tsallis statistics in a blast wave description,"
{\it Physical Review C} 79, no. 5 (2009): 051901(R),
https://doi.org/10.1103/PhysRevC.79.051901.

\bibitem{47i}
L. G. Gutay, A. S. Hirsch, C. Pajares, R. P. Scharenberg, and B.
K. Srivastava, ``De-confinement in small systems: clustering of
color sources in high multiplicity $\bar pp$ collisions at
$\sqrt{s}=1.8$ TeV," {\it International Journal of Modern Physics
E} 24, no. 12 (2015): 1550101,
https://doi.org/10.1142/S0218301315501013.

\bibitem{47j}
R. P. Scharenberg, B. K. Srivastava, and C. Pajares, ``Exploring
the initial stage of high multiplicity proton-proton collisions by
determining the initial temperature of the quark-gluon plasma,"
{\it Physical Review D} 100, no. 11 (2019): 114040,
https://doi.org/10.1103/PHYSREVD.100.114040.

\bibitem{47k}
P. Sahoo, S. De, S. K. Tiwari, and R. Sahoo, ``Energy and
centrality dependent study of deconfinement phase transition in a
color string percolation approach at RHIC energies," {\it The
European Physical Journal A} 54, no. 8 (2018): 136,
https://doi.org/10.1140/epja/i2018-12571-9.

\bibitem{47l}
Q. Wang and F. H. Liu, `Excitation function of initial temperature
of heavy flavor quarkonium emission source in high energy
collisions," {\it Advances in High Energy Physics} 2020 (2020):
5031494, https://doi.org/10.1155/2020/5031494.

\bibitem{47m}
Q. Wang, F. H. Liu, and K. K. Olimov, ``Initial-state temperature
of light meson emission source from squared momentum transfer
spectra in high-energy collisions," {\it Fronters in Physics
(Lausanne)} 9 (2021): 792039,
https://doi.org/10.3389/fphy.2021.792039.

\bibitem{47n}
M. Waqas and F. H. Liu, ``Initial, effective, and kinetic
freeze-out temperatures from transverse momentum spectra in high
energy proton(deuteron)-nucleus and nucleus-nucleus collisions,"
{\it The European Physical Journal Plus} 135, no. 2 (2020): 147,
https://doi.org/10.1140/epjp/s13360-020-00213-1.

\bibitem{47o}
D. K. Srivastava, R. Chatterjee, and M. G. Mustafa, ``Initial
temperature and extent of chemical equilibration of partons in
relativistic collision of heavy nuclei," arXiv:1609.06496,
https://doi.org/10.48550/arXiv.1609.06496.

\bibitem{47p}
R. A. Soltz, I. Garishvili, M. Cheng, B. Abelev, A. Glenn, J.
Newby, L. A. L. Levy, and S. Pratt, ``Constraining the initial
temperature and shear viscosity in a hybrid hydrodynamic model of
$\sqrt{s_{NN}}=200$ GeV Au+Au collisions using pion spectra,
elliptic flow, and femtoscopic radii," {\it Physical Review C} 87,
no. 4 (2013): 044901, https://doi.org/10.1103/PhysRevC.87.044901.

\bibitem{47q}
M. Csan{\'a}d, ``Direct photon spectra, flow and correlations from
hydro and implications on the initial temperature and EoS," {\it
Proceedings of Science} 154(WPCF2011) (2011): 035,
https://doi.org/10.22323/1.154.0035.

\bibitem{47r}
M. Csan{\'a}d, ``Initial temperature of the strongly interacting
quark gluon plasma created at RHIC," in: {\it Gribov-80 Memorial
Volume} (World Scientific, Singapore) (2011): 319--330,
arXiv:1101.1282, https://doi.org/10.1142/9789814350198\_0030.

\bibitem{47s}
M. Csan{\'a}d and I. M{\'a}jer, ``Initial temperature and EoS of
quark matter from direct photons," {\it Physics of Particles and
Nuclei Letters} 8, no. 9 (2011) 1013--1015,
https://doi.org/10.1134/S1547477111090147.

\bibitem{47t}
M. Csan{\'a}d and I. M{\'a}jer, ``Equation of state and initial
temperature of quark gluon plasma at RHIC," {\it Central European
Journal of Physics} 10, no. 4 (2012): 850--857,
https://doi.org/10.2478/s11534-012-0060-9.

\bibitem{47u}
F. Karsch, ``Lattice QCD at finite temperature and density," {\it
Nuclear Physics B -- Proceedings Supplements} 83--84 (2000):
14--23, https://doi.org/10.1016/S0920-5632(00)91591-3.

\bibitem{48}
D. Gallimore and J. F. Liao, ``A potential model study of the
nucleon's charge and mass radius," {\it Nuclear Physics A} 1055
(2025): 123012, https://doi.org/10.1016/j.nuclphysa.2024.123012.

\bibitem{49}
K. A. Bugaev, A. I. Ivanytskyi, V. V. Sagun, D. E. Grinyuk, D. O.
Savchenko, G. M. Zinovjev, E. G. Nikonov, L. V. Bravina, E. E.
Zabrodin, D. B. Blaschke, A. V. Taranenko, and L. Turko,
``Hard-core radius of nucleons within the induced surface tension
approach," {\it Universe} 5, no. 2 (2019): 63,
https://doi.org/10.3390/universe5020063.

\bibitem{50}
Z. Y. Zhu and A. Li, ``Nucleon radius effects on neutron stars in
quark mean field model," {\it Physical Review C} 97, no. 3 (2018):
035805, https://doi.org/10.1103/PhysRevC.97.035805.

\bibitem{50a}
F. Karsch, M. T. Mehr, and H. Satz, ``Color screening and
deconfinement for bound states of heavy quarks," {\it Zeitschrift
F{\"u}r Physik C} 37, no. 4 (1988): 617--622,
https://doi.org/10.1007/BF01549722.

\bibitem{50b}
B. Liu, P. N. Shen, and H. C. Chiang, ``Heavy quarkonium spectra
and $J/\psi$ dissociation in hot and dense matter," {\it Physical
Review C} 55, no. 6 (1997): 3021--3025,
https://doi.org/10.1103/PhysRevC.55.3021.

\bibitem{50c}
T. Das, ``Treatment of N-dimensional Schr{\"o}dinger equation for
anharmonic potential via Laplace transform," {\it Electronic
Journal of Theoretical Physics} 35 (2016): 207--214,
arXiv:1408.6139, https://doi.org/10.48550/arXiv.1408.6139.

\bibitem{50d}
T. Das, D. K. Choudhury, and K. K. Pathak, ``RMS and charge radii
in a potential model," {\it Indian Journal of Phys.} 90, no. 11
(2016): 1307--1312, https://doi.org/10.1007/s12648-016-0866-1.

\bibitem{50e}
G. M. Garc{\'i}a (for the ALICE Collaboration), ``Quarkonium
production measurements with the ALICE detector at the LHC," {\it
Journal of Physics G} 38, no. 12 (2011): 124034,
https://doi.org/10.1088/0954-3899/38/12/124034.

\bibitem{50f}
R. R. Ma (for the STAR Collaboration), ``Measurement of $J/\psi$
production in p+p collisions at $\sqrt{s}=500$ GeV at STAR
experiment," {\it Nuclear and Particle Physics Proceedings}
276--278 (2016): 261--264,
https://doi.org/10.1016/j.nuclphysbps.2016.05.059.

\bibitem{50g}
D. d'Enterria (for the CMS Collaboration), ``High-density QCD with
CMS at the LHC," {\it Journal of Physics G} 35, no. 10 (2008):
104039, https://doi.org/10.1088/0954-3899/35/10/104039.

\bibitem{50h}
J. Kim, J. Seo, B. Hong, J. Hong, E. J. Kim, Y. Kim, M. Kweon, S.
H. Lee, S. Lim, and J. Park, ``Model study on $\Upsilon(nS)$
modification in small collision systems," {\it Physical Review C}
107, no. 5 (2023): 054905,
https://doi.org/10.1103/PhysRevC.107.054905.

\bibitem{51}
Y. Q. Gao and F. H. Liu, ``Comparing Tsallis and Boltzmann
temperatures from relativistic heavy ion collider and large hadron
collider heavy-ion data," {\it Indian Journal of Physics} 90, no.
3 (2016): 319--334, https://doi.org/10.1007/s12648-015-0747-z.

\bibitem{52}
L. N. Gao, F. H. Liu, and R. A. Lacey, ``Excitation functions of
parameters in Erlang distribution, Schwinger mechanism, and
Tsallis statistics in RHIC BES program," {\it The European
Physical Journal A} 52, no. 5 (2016) 137,
https://doi.org/10.1140/epja/i2016-16137-7.

\bibitem{53}
L. N. Gao and F. H. Liu, ``Comparing Erlang distribution and
Schwinger mechanism on transverse momentum spectra in high energy
collisions," {\it Advances in High Energy Physics} 2016 (2016):
1505823, http://dx.doi.org/10.1155/2016/1505823.

\bibitem{54}
T. T. Duan, P. P. Yang, P. C. Zhang, H. L. Lao, F. H. Liu, and K.
K. Olimov, ``Comparing effective temperatures in standard,
Tsallis, and q-dual statistics from transverse momentum spectra of
identified light charged hadrons produced in gold--gold collisions
at RHIC energies," {\it The European Physical Journal Plus} 139,
no. 12 (2024): 1069,
https://doi.org/10.1140/epjp/s13360-024-05853-1.

\bibitem{55}
P. C. Zhang, P. P. Yang, T. T. Duan, H. L. Zhu, F. H. Liu, and K.
K. Olimov, ``Comparing effective temperatures in standard and
Tsallis distributions from transverse momentum spectra in small
collision systems," {\it Indian Journal of Physics} (2025) online
first, https://doi.org/10.1007/s12648-025-03742-6.

\bibitem{67a}
A. Bialas, ``Fluctuations of the string tension and transverse
mass distribution," {\it Physics Letters B} 466, nos. 2--4 (1999):
301--304, https://doi.org/10.1016/S0370-2693(99)01159-4.

\bibitem{67b}
W. Florkowski, ``Schwinger tunneling and thermal character of
hadron spectra," {\it Acta Physica Polonica B} 35, no. 2 (2004):
799--807,
https://www.actaphys.uj.edu.pl/fulltext?series=Reg\&vol=35\&page=799.

\bibitem{67c}
N. Zardoshti (for the ALICE Collaboration), ``First direct
observation of the dead-cone effect," {\it Nuclear Physics A} 1005
(2021) 121905, https://doi.org/10.1016/j.nuclphysa.2020.121905.

\bibitem{67d}
ALICE Collaboration (S. Acharya {\it et al.}), ``Direct
observation of the dead-cone effect in QCD," {\it Nature} 605, no.
7910 (2022): 440--446, https://doi.org/10.1038/s41586-022-04572-w.

\bibitem{67e}
R. Thomas, B. Kampfer, and G. Soff, ``Gluon emission of heavy
quarks: dead cone effect," {\it Acta Physica Hungarica Series A}
22, no. 1--2 (2005): 83--91,
https://doi.org/10.1556/APH.22.2005.1-2.9.

\bibitem{67f}
D. R. Herrera, J. R. A. Garc{\'i}a, A. F. T{\'e}llez, J. E.
Ram{\'i}rez, and C. Pajares, ``Entropy and heat capacity of the
transverse momentum distribution for pp collisions at RHIC and LHC
energies," {\it Physical Review C} 109, no. 3 (2024): 034915,
https://doi.org/10.1103/PhysRevC.109.034915.

\bibitem{67g}
H. J. Pirner, B. Z. Kopeliovich, and K. Reygers, ``Strangeness
enhancement due to string fluctuations," {\it Physical Review D}
101, no. 11 (2020): 114010,
https://doi.org/10.1103/PhysRevD.101.114010.

\bibitem{68}
R. N. Patra (for the ALICE Collaboration), ``Collective phenomena
study in small systems using the bulk of particle production in
high-multiplicity pp collisions at $\sqrt{s}=13$ TeV with ALICE,"
{\it Proceedings of Science} 449(EPS-HEP2023) (2023): 214,
https://doi.org/10.22323/1.449.0214.

\bibitem{69}
J. F. Grosse-Oetringhaus and U. A. Wiedemann, ``A decade of
collectivity in small systems," {\it World Scientific Annual
Review of Particle Physics} (2025) Invited article submitted,
arXiv:2407.07484, https://doi.org/10.48550/arXiv.2407.07484.

\bibitem{70}
T. Song, J. Aichelin, and E. Bratkovskaya, ``The production of
primordial $J/\psi$ in p+p and relativistic heavy-ion collisions,"
{\it Physical Review C} 96, no. 1 (2017): 014907,
https://doi.org/10.1103/PhysRevC.96.014907.

\bibitem{71}
B. Paul, M. Mandal, P. Roy, and S. Chattapadhyay, ``Systematic
study of charmonium production in pp collisions at the LHC
energies," {\it Journal of Physics G} 42, no. 6 (2015): 065101,
https://doi.org/10.1088/0954-3899/42/6/065101.

\bibitem{72}
C. Tsallis, ``Possible generalization of Boltzmann--Gibbs
statistics," {\it Journal of Statistical Physics} 52, nos. 1--2
(1988): 479--487, https://doi.org/10.1007/BF01016429.

\bibitem{73}
C. Tsallis, ``Nonadditive entropy and nonextensive statistical
mechanics---an overview after 20 years," {\it Brazilian Journal of
Physics} 39, no. 2a (2009): 337--356,
https://doi.org/10.1590/S0103-97332009000400002.

\bibitem{74}
T. S. Bir{\'o}, G. Purcsel, and K. Urm{\"o}ssy, ``Non-extensive
approach to quark matter," {\it The European Physical Journal A}
40, no. 3 (2009): 325--340,
https://doi.org/10.1140/epja/i2009-10806-6.

\bibitem{75}
J. Cleymans and M. W. Paradza, ``Statistical approaches to high
energy physics: chemical and thermal freeze-outs," {\it Physics}
2, no. 4 (2020): 654--664, https://doi.org/10.3390/physics2040038.

\bibitem{76}
A. S. Parvan, ``Equivalence of the phenomenological Tsallis
distribution to the transverse momentum distribution of q-dual
statistics," {\it The European Physical Journal A} 56, no. 4
(2020): 106, https://doi.org/10.1140/epja/s10050-020-00117-9.

\bibitem{77}
N. Fischer and T. Sj{\"o}strand, ``Thermodynamical string
fragmentation," {\it Journal of High Energy Physics} 2017, no. 1
(2017): 140, https://doi.org/10.1007/JHEP01(2017)140.

\bibitem{78}
H. J. Pirner, B. Z. Kopeliovich, and K. Reygers, ``Strangeness
enhancement due to string fluctuations," {\it Physical Review D}
101, no. 11 (2020): 114010,
https://doi.org/10.1103/PhysRevD.101.114010.

\bibitem{79}
J. R. A. Garc{\'i}a, D. R. Herrera, P. Fierro, J. E. Ram{\'i}rez,
A. F. T{\'e}llez, and C. Pajares, ``Soft and hard scales of the
transverse momentum distribution in the color string percolation
model," {\it Journal of Physics G} 50, no. 12 (2023): 125105,
https://doi.org/10.1088/1361-6471/acffe1.

\bibitem{80}
D. R. Herrera, J. R. A. Garc{\'i}a, A. F. T{\'e}llez, J. E.
Ram{\'i}rez, and C. Pajares, ``Nonextensivity and temperature
fluctuations of the Higgs boson production," {\it Physical Review
C} 110, no. 1 (2024): 015205,
https://doi.org/10.1103/PhysRevC.110.015205.

\bibitem{81}
H. Boschi-Filho, N. R. F. Braga, and C. N. Ferreira, ``Heavy quark
potential at finite temperature from gauge/string duality," {\it
Physical Review D} 74, no. 8 (2006): 086001,
https://doi.org/10.1103/PhysRevD.74.086001.

\bibitem{82}
V. Mateu, P. G. Ortega, D. R. Entem, and F. Fernandez,
``Calibrating the na{\"i}ve Cornell model with NRQCD," {\it The
European Physical Journal C} 79, no. 4 (2019): 323,
https://doi.org/10.1140/epjc/s10052-019-6808-2.

\end{thebibliography}}
\end{document}